\documentclass[runningheads]{llncs}

\usepackage{amssymb}
\usepackage{enumitem}
\usepackage{wrapfig}

\usepackage{graphicx}
\usepackage{xspace}
\usepackage{amsmath}
\usepackage{tablefootnote}
\usepackage[leftcaption]{sidecap}
\usepackage{color}
\definecolor{colLink}{rgb}{0.0,0,0.7}
\definecolor{colCite}{rgb}{0.0,0,0.7}
\definecolor{colURL}{rgb}{0,0,0.7}

\usepackage{hyperref}

\usepackage{multirow}
\usepackage{url}

\newcommand{\comments}[1]{}

\newcommand{\eg}{\emph{e.g.}}

\newcommand{\hide}[1]{}

\newcommand{\algformat}[1]{\textsc{#1}\xspace}
\newcommand{\minisat}{\algformat{MiniSat}}
\newcommand{\monosat}{\algformat{MonoSat}}
\newcommand{\clasp}{\algformat{clasp}}

\newcommand{\gunder}{G_{under}}
\newcommand{\gover}{G_{over}}

\newlength{\lenLNCSFigureSquareTwo}
\newcommand{\LNCSFigureSquareTwo}[4]{
	\begin{figure}[!tb]
	\centering
\mbox{\includegraphics[width=\lenLNCSFigureSquareTwo] {#1}}
~~~~
\mbox{\includegraphics[width=\lenLNCSFigureSquareTwo] {#2}}
	\caption[#3]{\textbf{#3}.  #4}
	\label{#1}
	\end{figure}
}

\begin{document}

\mainmatter  


\title{SAT Modulo Monotonic Theories}
\titlerunning{SAT Modulo Monotonic Theories}


\author{Sam Bayless\inst{1}, Noah Bayless\inst{2}, Holger~H.~Hoos\inst{1}, Alan~J.~Hu\inst{1}}
\authorrunning{Bayless, Bayless, Hoos, and Hu}
\tocauthor{Sam Bayless, Noah Bayless, Holger~H.~Hoos, Alan~J.~Hu}
\institute{Dept. of Computer Science, Univ. of British Columbia, Canada\\ \email{\{sbayless,hoos,ajh}@cs.ubc.ca\}\\ \and
 Point Grey Mini Secondary School, Vancouver, Canada\\ \email{nbayless@pgmini.org}}

\maketitle

\begin{abstract}

We define the concept of a ``monotonic theory'' and show how to build
efficient SMT (SAT Modulo Theory) solvers, including effective theory
propagation and clause learning, for such theories.
We present examples showing that monotonic theories
arise from many common problems, e.g., graph properties
such as reachability, shortest paths, connected components,
minimum spanning tree, and max-flow/min-cut, and then demonstrate
our framework by building SMT solvers for each of these theories.
We apply these solvers to procedural content generation
problems, demonstrating major speed-ups over state-of-the-art
approaches based on SAT or Answer Set Programming, and easily
solving several instances that were previously impractical to solve.
\end{abstract}


\section{Introduction}\label{intro}

One contributing reason for the success of Boolean satisfiability (SAT) solvers has been the use of SAT Modulo Theory solvers to extend SAT to domains that would otherwise be impractical or impossible to represent or solve in SAT.  However, designing efficient SMT solvers typically requires expertise and deep insight for each new theory.

In this work,  we identify a class of theories for which we can create
efficient SMT solvers either partially or entirely automatically. These
theories --- which we term `monotonic'\footnote{To forestall confusion,
note that our concept of a `monotonic theory' here has no direct
relationship to the concept of monotonic/non-monotonic logics.} ---
share
characteristics that admit straightforward techniques
for building lazy SMT theory solvers.
We show that many common problems
can be tackled using these techniques, and demonstrate this in practice
by building efficient SMT solvers for various graph properties:
reachability, shortest paths, connected components, minimum spanning
trees, and minimum-cut/maximum-flow. These graph properties are useful
for solving many content generation tasks, such as maze and
terrain generation.
We also observe that the optima of constrained optimization problems
are monotonic with respect to their constraints, making our work
applicable to a wide variety of theories.
We illustrate this with a simple job-scheduling theory.

We implement solvers for these theories in our tool \monosat,
based on \minisat~2.2~\cite{een2004extensible}, and apply it
to realistic declarative procedural content generation tasks,
showing large speed-ups over the Answer Set Programming (ASP)
solver~\clasp\cite{gebser2007clasp}, the state of the art for these tasks. We
exhibit dramatic speed-ups over both \minisat and \clasp, solving many instances that both are unable to solve
in a reasonable amount of time. 

It has recently come to our attention that some or all of the ASP encodings we compare to 
in this paper are sub-optimal; in fact, we have already seen improvements on some but not all of the \clasp results (as well as improvements in our own solver's results). We include these experimental results in the following, but intend to replace them in the final version 
of this study.

\section{Monotonic Theories\label{sec:smmt}}

%

We define a Boolean monotonic predicate as:

\begin{definition}[Monotonic Predicate]
A predicate $p$ : $\{0,1\}^n \mapsto \{0,1\}$ is (Boolean) monotonic if, and only if,  for all $s_i$, the following property holds:
\begin{eqnarray}
\label{condition:predicate}   p(\ldots, s_{i-1}, 0, s_{i+1} \ldots) & \rightarrow & p(\ldots, s_{i-1}, 1, s_{i+1} \ldots) 
\end{eqnarray}
\end{definition}

We have given the definition for a positive monotonic predicate; a symmetric definition exists for the negative case. Notice that the domain of $p$ is restricted to Booleans; monotonic functions over the Booleans are also known as \textit{unate} functions. 

The notion of a monotonic predicate has been previously explored in~\cite{bradley2007checking,bradley2008property,marques2013minimal}, in the context of finding a minimal models and unsatisfying subsets of a formula. Our presentation differs slightly from the one given in~\cite{bradley2007checking,bradley2008property,marques2013minimal}, who define a monotonic predicate as $p: 2^S \mapsto {0,1}$, over the powerset of some set $S$, such that: 

\begin{definition}[Monotonic Predicate (Set-Oriented)]
Given a set $S$, a predicate $p$ : $2^S \mapsto \{0,1\}$ is monotonic if, and only if:
\begin{eqnarray}
p(S) \text{ holds.}\\
\text{If } p(S_0) \text{ holds and } S_0 \subseteq S_1 \subseteq S \text{ then } p(S_1) \text{ holds.}
\end{eqnarray}
\end{definition}

We relax the condition that p(S) must hold (so that the constant function $p() = 0$ can be considered monotonic). The second condition, using the common mapping from bit-vectors to set inclusions, is equivalent to our definition (\ref{condition:predicate}). We now define a notion of a monotonic theory in the context of SMT:

\begin{definition}[Monotonic Theory]
A theory $T$ with signature $\Sigma$ is a Boolean monotonic theory if, and only if:
\begin{enumerate}
\item\label{condition:boolean}   The only sort in $\Sigma$ is Boolean.
\item\label{condition:mono}   All predicates  in $\Sigma$ are monotonic.   
\item\label{condition:function}   All functions in $\Sigma$ are monotonic.\footnote{As all sorts are Booleans, all functions are technically predicates. However, SMT formulations typically make a distinction between top-level predicates in a formula (which instantiate atoms) and functions as arguments to predicates or other functions (which instantiate terms), and we respect that distinction in this paper.}  
\end{enumerate}
\end{definition}

We will allow both positive and negative monotonic functions (and predicates). As is typical for SMT solvers, we consider only decidable,
quantifier-free, first-order theories. Atypically, the sorts are restricted to Booleans; however, we will show that even such a limited theory can express useful predicates. We note that although any (decidable) predicate over the domain of Booleans can in principle be encoded into CNF, there are many such functions for which we don't know of any efficient encodings, or for which the best known encoding may require a super-linear number of constraints. It is these functions that we are interested in solving efficiently in this paper.

\comments{
Finally, because the truth value of each predicate is defined as a function over the atoms of predicates in $S$, and $S \cap P = \emptyset$, we also have that for any $P$-atoms $p$ and $q$, and for any complete truth assignment $M_S$ to the $S$-atoms:
\begin{eqnarray}
\label{condition:independence_under} SAT_T(M_S \cup  \{p\}) \land SAT_T(M_S \cup  \{q\})  & \rightarrow  & SAT_T(M_S \cup  \{p, q\})\\
\label{condition:independence_over} SAT_T(M_S \cup  \{ \lnot p\}) \land SAT_T(M_S \cup  \{ \lnot q\})  & \rightarrow  & SAT_T(M_S \cup  \{ \lnot p, \lnot q\})
\end{eqnarray}
}

We will assume in this paper that each predicate $p$ is positive monotonic; a symmetric definition can be given for the negative case, and in general we are free to invert the semantics of individual predicates to make either form fit, as needed. 

As an illustrative example, consider a theory of graph reachability,
with predicates of the form $reach_{u,v,G}(edge_1, edge_2, edge_3,\ldots)$, where each $edge_i$ is a Boolean variable.
Notice that $reach_{u,v,G}$ describes a family of predicates over the edges of graph $G$: 
for each pair of vertices $u,v$, and for each graph $G=(V,E)$ with $|V|$ vertices
and $|E|$ edges that \textit{may} or \textit{may not} be included in the graph, there is a separate
$reach$ predicate in the theory. The Boolean arguments $edge_i$ 
define which edges (via some mapping to the fixed set of possible edges $E$) are included in the graph.
$reach_{u,v,G}$ is monotonic with respect to the set of edges in a graph:
if a node $v$ is reachable from another node $u$ in a given graph $G$ that does not contain
$edge_i$, then it must still be reachable in an identical graph that also contains $edge_i$.
Conversely, if a node $v$ is not reachable from node $u$ in graph $G$, then removing an edge from $G$ 
cannot make $v$ reachable.

\section{Theory Propagation and Clause Learning\label{sec:propagation}}


Many successful SMT solvers follow the \textit{lazy} SMT
design~\cite{sebastiani2007lazy,de2008z3}, in which a SAT solver is
combined with a set of theory solvers, and each theory solver supplies
(at least) two capabilities:  (1) theory propagation (or
theory deduction), which takes a partial assignment $M$ to the theory
atoms for that theory, and checks if any other atoms are implied by
that partial assignment (or if $M$ constitutes a conflict in the theory
solver), and (2) clause learning (equivalently, deriving conflict
or justification sets), where given a conflict $c$ in $M$, the theory
solver finds a subset of $M$ sufficient to imply $c$, which the SAT
solver can then negate and store as a learned clause. The effectiveness
of a lazy SMT theory solver depends on the ability of the theory solver
to propagate atoms and detect conflicts early, from small assignments $M$, and to
find small conflict sets in $M$ when a conflict does arise.

Many theories have known, efficient
algorithms for deciding the satisfiability of \textit{fully} specified inputs, but not for partially
specified or symbolic inputs. Continuing with the reachability example from above, given a concretely specified graph, one can find the set of nodes reachable from $u$ simply using DFS. In contrast, determining whether a node is reachable in a graph with symbolically defined edges is less obvious. Given only a procedure for computing the truth-values of the monotonic predicates
$p$ from complete assignments, we will show how we can take advantage of the properties of a monotonic theory
to form a complete and efficient decision procedure for any Boolean monotonic theory. 

Given a formula $\phi$ in the language of some monotonic theory, we first apply the following two transformations, both requiring
linear time: first, for each function symbol $f$ in the theory language, we introduce a new predicate $p_f$ whose truth-value is given by the monotonic function $f$.
We then replace all occurrences of any term $f(a,b,c,\ldots)$ with a new Boolean variable, $v_f$, and conjoin the constraint $(v_f \Leftrightarrow p_f(a,b,c,\ldots))$ to $\phi$. For simplicity, we will assume in the following that we have also applied the same transformation to occurrences of negated terms (that is, we treat logical negation as a negative monotonic function, and replace all negated terms (but \textit{not} atoms) with logically equivalent non-negated Boolean variables), but in practice this can be easily avoided in the solver.
Secondly, let $expose(v)$ be a trivial positive monotonic predicate of arity 1 that is true exactly when Boolean variable $v$ is true. 
We now expose each Boolean variable $v$ in $\phi$ as an atom in the search space of the SAT solver, by conjoining the clause $(expose(v) \lor \lnot expose(v))$ to $\phi$. We then collect these exposed atoms in the set $S$ and all other atoms in the set $P$. As a result, for any atom $p \in P$, exposed atoms equivalent to the Boolean arguments of atom $p$ can be found in $S$.

Observe that as all the newly introduced predicates are monotonic predicates, and all introduced variables are Boolean, this transformed theory is still a monotonic theory. In this transformed formula, all terms are Boolean variables, as all functions have been replaced with logically equivalent predicates. Without loss of generality, we will assume that all predicates are positive monotonic predicates; a simple transformation can ensure this is the case.

This simplified, exposed formula now has the following useful properties: for any atom $p \in P$, and any partial truth assignment $M_S$ to the atoms of the exposed Boolean variables $S$ \textit{excluding} variable $s \in S$,
\begin{eqnarray}
\label{condition:under}   SAT_T (M_{S} \cup \{\lnot s, p\})  & \rightarrow  &   SAT_T (M_{S} \cup \{s, p\})\\
\label{condition:over}    SAT_T (M_{S} \cup \{s, \lnot p\})  & \rightarrow  &   SAT_T (M_{S} \cup \{\lnot s, \lnot p\})
\end{eqnarray}




Given a (partial) truth assignment $M$, let $M_S$ be corresponding (partial) assignment to just the $S$-atoms. We form two
completions of $M_S$: one in which all the unassigned $S$-atoms
are assigned to false ($M_S^-$),
and one in which they are assigned to true ($M_S^+$). Since $M_S^-$ contains
a complete assignment to the $S$-atoms, we can use it as input to any standard algorithm
for computing the truth-value of atom $p \in P$ from concrete inputs, which will determine
whether $M_S^- \implies p$. By property (\ref{condition:under}), if $M_S^-
\implies p$, then $M_S \implies p$.  Similarly,
by property (\ref{condition:over}), if
$M_S^+ \implies \lnot p$, then $M_S \implies \lnot p$. If either case holds,
we can propagate $p$ (or $\lnot p$) back to the SAT solver. Thus,
$M_S^-$ and $M_S^+$ allow us to safely under- and over-approximate the truth value of
$p$. We can apply this technique iteratively for each $P$-atom.

This over/under-approximation scheme gives us theory propagation for $P$-atoms \textit{only};
however, because there can be no conflicts among the atoms of $S$ by themselves, applying propagation to the $P$-atoms is 
sufficient to detect any conflicting assignment. Furthermore, because all variables in the formula are exposed to the SAT solver's search space,
this technique provides a complete procedure for deciding the satisfiability of any monotonic theory $T$,
so long as procedures are available for computing each monotonic predicate from complete assignments to their arguments.
As we will show later,
this is also sufficient to build efficient solvers for a wide set of
theories. Most importantly, it allows us to use \textit{standard}
algorithms for solving the concrete, fully specified and non-symbolic
forms of these theories, 
and without any modification, to apply theory propagation to $P$.
Section~\ref{sec:theories}
introduces several theory solvers based on this framework
(e.g., using Dijkstra's algorithm for shortest paths);
no special data structures or modifications to any of these
algorithms were required to obtain efficient theory propagation.\footnote{One observation is that this scheme \textit{does} require the theory solver
to know which atoms are unassigned in the formula, which is information that the
standard Lazy-SMT framework doesn't expose to theory solvers. 
One general way to resolve this would be to explicitly pass a set of unassigned
$S$-atoms, $S_{unassigned}$, along with the partial assignment $M$, to the theory
propagation routine. In the theories we will present in this
paper, the theory solver will always have enough information on its own to reconstruct
the unassigned $S$-atoms from a partial assignment $M$; for example, in the 
theory of graph reachability, each $edge$ atom takes as an argument a 
graph object, which contains a set $E$ of all edges (assigned or not); 
by comparing the assigned edges to $E$, the unassigned edges are easily deduced.}

What about clause learning? In general, given a conflict on
a partial assignment $M$ over the theory's atoms, a theory solver
can always simply return the clause $\lnot M$, which states that at
least one of the assigned theory atoms must change. However,
for conflicts arising from the over/under-approximation
theory-propagation scheme above, we can do better. First, by property (\ref{condition:mono}),
all assignments to just the $S$-atoms are satisfiable, so any conflicting
assignment must assign at least one $P$-atom.

Secondly, observe that the over/under-approximations scheme only computes implications
from assignments of the $S$-atoms $M_S$ to individual $P$-atoms, and never computes
implications from, for example, $P$-atoms to $S$-atoms or from $P$-atoms to other $P$-atoms.
For this reason, in any conflict discovered by this scheme, there must exists a \textit{single}
$P$-atom $p$, such that the assignment to the $S$-atoms, along with the assignment to $p$, are together UNSAT.
Note that this may not hold if other techniques are used to apply theory propagation from, for
example, the $P$-atoms to the $S$-atoms.

By property
(\ref{condition:under}), if $M_S^- \implies p$, then the
positive theory atoms in $M_S$ are sufficient to imply $p$ by themselves.
By property (\ref{condition:over}), if $M_S^+ \implies \lnot p$, then the
negative theory atoms in $M_S$ are sufficient to imply $\lnot p$ by
themselves. Therefore, the positive (resp. negative) assignments to the atoms in $M_S$ alone
can safely form justification sets for $p$ (resp. $\lnot p$).
This is already an improvement over simply learning the clause $\lnot M$,
but in many cases we can do even better.

Many common algorithms are constructive in the sense that they not only
compute whether $p$ is true or false, but also produce a witness (in terms
of the inputs of the algorithm) that is a sufficient condition to imply
that property. In many cases, the witness will be constructed strictly
in terms of inputs corresponding to atoms that are assigned to true (or
alternatively, strictly in terms of atoms that are assigned false). This
need not be the case --- the algorithm might not be constructive,
or it might construct a witness in terms of some combination of the
true and false atoms in the assignment --- but, as we will show below,
it commonly \textit{is} the case for many theories of interest.  For example,
if we used breadth-first search to find that node $v$ is reachable
from node $u$ in some graph,
then we obtain as a side effect a path from $u$
to $v$, and the theory atoms corresponding to the edges in that path
imply that $v$ can be reached from $u$.

Any algorithm that can produce a witness containing solely
positive $S$-atoms can be used to produce justification sets
(or learned clauses) for any positive $P$-atom assignments propagated
by the scheme under-approximation above. Similarly, any algorithm that
can produce a witness of negative $S$-atoms can also produce
justification sets for any negative $P$-atom assignments propagated by
over-approximation.
Some algorithms can produce both, but this isn't guaranteed to be the case.
In practice, we have often found standard algorithms that produce
one, but not both, types of witnesses.
For the missing witness, we can always safely fall back on the
strategy of using the positive (resp.~negative) theory atoms of $M_S$
as the justification sets for $p$ (resp.~$\lnot p$).

\section{Equality Relations in Monotonic Theories\label{sec:equality}}

As equality relations (over non-trivial domains) are non-monotonic, monotonic theories cannot include equality predicates.
For this reason, none of the theories that we will describe in this paper support equality relations. This is fairly unusual in SMT; in fact, in many definitions of SMT\cite{sebastiani2007lazy,de2009satisfiability}, the equality relation is treated as a \textit{built-in} operator that is implicitly available to all theories.

However, as all sorts are restricted to Booleans, and equality semantics over Booleans can be enforced outside the theory solver (in the CNF formula), this limitation does not limit the expressivity of Boolean monotonic theories. We further note that theory combination for (disjoint) Boolean monotonic theories is trivial (once they have been transformed as in \autoref{sec:propagation}), as they are function-free and restricted to Booleans. 

\section{Examples of Monotonic Theory Solvers \label{sec:theories}}

We now introduce several monotonic theories and show how effective SMT
solvers can be built for each of them using the theory propagation and
clause learning strategies from Section~\ref{sec:propagation}.  First,
we'll consider several common graph properties.
Afterward, we will illustrate via a simple job-scheduling theory
how monotonic theories naturally arise from
typical constraint satisfaction problems.
Section~\ref{sec:graph} shows applications of these theories
and presents results.

In each of our graph theories, we have a finite set of vertices $V$
and edges $E \subseteq V \times V$,
and the solver needs to select a subset of edges to include in a graph $G$.
Each edge $(u,v) \in E$ is associated with a Boolean variable, exposed to the SAT solver by theory
atom $edge_{u,v,G}$, and is in graph $G$ if and only if
$edge_{u,v,G}$ is true (this definition can be easily extended to
support weighted edges and multigraphs, if desired). Each theory below
also supports a monotonic predicate for $G$; for example, the graph
reachability theory has predicates of the form $reach_{u,v,G}(edges)$, which are
true if and only if node $v$ is reachable from node $u$ in graph $G$,
under a given assignment to the \textit{edge} variables, in the vector $edges$. We will speak of
edges being ``enabled'' or ``disabled'' as shorthand for the truth values of
these $edge$ variables and their corresponding atoms. The edge atoms form the set $S$ (as defined in \autoref{sec:propagation}), while the various graph predicates we consider will form sets $P$.

As previously observed, given a graph (directed or undirected) and some fixed starting node $u$, adding an edge can increase the set of nodes that are reachable from $u$, but cannot decrease it.  Removing an edge from the graph can decrease the set of nodes reachable from $u$, but cannot increase it. The other graph properties we consider are monotonic with respect to the edges in the graph in same way; for example, adding an edge can decrease the weight of the minimum spanning tree, but not increase it. 

For each solver, given partial truth assignment $M$, we will construct two
auxiliary graphs, $\gunder$ and $\gover$. The graph $\gunder$ is formed
from the edge assignments in $M_S^-$: only edges that are assigned true in
$M$ are included in $\gunder$; edges that are either assigned to false
or are unassigned in $M$ are excluded. In our second graph, $\gover$, we
include all edges that are either assigned to true \textit{or} unassigned
in $M$, corresponding to the edge assignments in $M_S^+$. We then
apply standard graph algorithms
to $\gunder$ and $\gover$ during theory propagation, using the
over/under-approximation scheme described in Section \ref{sec:smmt}.
Furthermore, because $\gunder$ and $\gover$ are completely specified
concrete graphs,
there exists a wide library of efficient graph algorithms that we could
apply for different types of graphs.
For example, we could
easily specialize the algorithm if we knew that the graphs were
undirected, acyclic, sparse, or planar. In the case of reachability
and shortest paths, we could use an all-pairs shortest-paths algorithm,
such as Floyd-Warshall, if we knew that many queries from different
source nodes were being made.

The over/under-approximation theory propagation scheme involves making
repeated graph queries for our theory solver --- one for each of $\gunder$
and $\gover$, for each new partial assignment generated by the SAT
solver. One improvement we did implement is to check if under the
current partial assignment, any new edges have been added to $\gunder$,
or removed from $\gover$, and only recompute our graph property for
the graph(s) that have changed since our last check.  This check can
be performed very efficiently, and we find that in many cases, most
of the time will either be spent assigning edges to true or assigning
edges to false, so this check often lets us skip roughly half of the
computations required otherwise. A further possible improvement (which
we have not implemented) would be to use a dynamic algorithm for the
graph property in question, that can be updated efficiently as edges
are added and removed from the graph.



\subsection{{Graph Reachability}}
The first SMT solver we consider is for graph reachability, which we showed to be monotonic above (see Section \ref{sec:smmt}).

\begin{description}
\item[Monotonic Predicate:] $reach_{u,v,G}(edges)$, true iff $u$ can reach $v$ in $G$ given $edges$. 
\item[Algorithm:] Breadth-first search (BFS).
\item[Conflict set for $reach_{u,v,G}(edges)$:] Let $e_1, e_2, \ldots$ be a $u-v$ path in $\gunder$; the conflict set is $\{\lnot e_1, \lnot e_2, \ldots, reach_{u,v,G}(edges)\}$. 

\item[Conflict set for $\lnot reach_{u,v,G}(edges)$:] Let  $e_1, e_2, \ldots$ be a cut of disabled edges separating $u$ from $v$ in $\gover$. The conflict set is $\{e_1, e_2, \ldots, \lnot reach_{u,v,G}(edges)\}$.

We can find the cut by setting the weight of each disabled edge in $M$
to 1, the weight of all other edges to infinity, and finding a minimal
$u-v$ cut. However, in our implementation, we found it more efficient to simply
walk back from $u$ in $\gover$, through the enabled and unassigned edges,
and to add each incident \textit{disabled} edge to the cut.

\item[Decision Heuristic:] (Optional) If $reach_{u,v,G}(edges)$ is assigned to be true in $M$, but there does not yet exist a $u-v$ path in $\gunder$,  then find a $u-v$ path in $\gover$ and pick the first unassigned edge in that path to be assigned true as the next decision. This heuristic is effective when the instance is dominated by one very difficult reachability constraint. 


\end{description}

\subsection{{Shortest Paths}}
Given a (possibly weighted, possibly directed) graph, the shortest path between nodes $u$ and $v$ can decrease in length as edges are added, but cannot increase. 

\begin{description}
\item[Monotonic Predicate:] $shortestPath_{u,v,G \leq C}(edges)$, true iff the shortest $u-v$ path $\leq C$ in $G$, given \textit{edges}. 
\item[Algorithm:] Dijkstra's Algorithm for Shortest Path~\cite{dijkstra1959note}.
\item[Conflict set for $shortestPath_{u,v,G \leq C}(edges)$:] Let $e_1, e_2, \ldots$
be the shortest $u-v$ path in $\gunder$.  The theory solver has determined
that the weight of this path $\le C$.  The conflict set is
then $\{\lnot e_1, \lnot e_2, \ldots, shortestPath_{u,v,G \leq C}(edges)\}$.
\item[Conflict set for $\lnot shortestPath_{u,v,G \leq C}(edges)$:] Walk back from $u$ in $\gover$ along enabled and unassigned edges; collect all incident disabled edges $e_1, e_2, \ldots$; the conflict set is $\{e_1, e_2, \ldots, \lnot shortestPath_{u,v,G \leq C}(edges)\}$.

\end{description}

\subsection{{Connected Components}}
The reachability solver above can be directly combined with a union-find data structure (either disjoint-sets, or a fully dynamic data structure\cite{holm2001poly}) to efficiently compute all-pairs connectivity in undirected graphs. We consider here a different question: determining whether the number of connected components in an unweighted, undirected graph is less than or equal to some constant. 

\begin{description}
\item[Monotonic Predicate:] $connectedComponents_{G \leq C}(edges)$, true iff the number of connected components in $G$ is $\leq C$, given \textit{edges}. 
\item[Algorithm:] Disjoint-Sets/Union-Find.
\item[Conflict set for $connectedComponents_{G \leq C}(edges)$:]  Construct a spanning tree for each component of $\gunder$ (we can do this simply using DFS). Let edges $e_1, e_2, \ldots$ be the edges in these spanning trees; the conflict set is\\ $\{\lnot e_1, \lnot e_2, \ldots, connectedComponents_{G \leq C}(edges)\}$.
\item[Conflict set for $\lnot connectedComponents_{G \leq C}(edges)$:] Collect all disabled edges $e_i=(u,v)$ where $u$ and $v$ belong to different components in $\gover$;
the conflict set is $\{e_1, e_2, \ldots, \lnot connectedComponents_{G \leq C}(edges)\}$.

\end{description}

\subsection{{Maximum-Flow/Minimum-Cut}}
Here we consider the maximum $s-t$ flow (or minimum
$s-t$ cut) in a weighted, directed graph. We extend the edges of $G$ to have associated weights. The
weights are assumed to be fixed constants for each edge (but the theory
can handle multiple possible weights on each edge by
supporting multiple edges with different weights between two nodes).

\begin{description}
\item[Monotonic Predicate:] $maxFlow_{s,t,G \geq C}(edges)$, true iff the maximum $s-t$ flow in $G$ is $\geq C$.
\item[Algorithm:] Edmonds-Karp's algorithm for maximum-flow/minimum-cut~\cite{edmonds1972theoretical}.
\item[Conflict set for $maxFlow_{s,t,G \geq C}(edges)$:]  If the maximum flow
in $\gunder$ is $\geq C$,
the conflict set is $\{\lnot e_1, \lnot e_2, \ldots, maxFlow_{s,t,G \geq C}(edges)\}$, where $\{e_1, e_2, \ldots\}$ are the edges with non-zero flow.
\item[Conflict set for $\lnot maxFlow_{s,t,G \geq C}(edges)$:] Find a cut
$\{e_1,e_2,\ldots\}$ of disabled edges in the residual $s-t$ flow graph
for $\gover$. As in the case of reachability, we could search for a
minimum cut, but we found it more effective to simply walk back
from $v$ to $u$ in the residual graph, collecting incident disabled edges;
the conflict set is $\{e_1, e_2, \ldots, \lnot maxFlow_{s,t,G \geq C}(edges)\}$.
\end{description}

\subsection{{Minimum Weight Spanning Trees}}
Here we consider constraints on the weight of the minimum spanning tree in a weighted (non-negative) undirected graph. The weights are assumed to be fixed constants for each edge, but there may be multiple edges with different weights between any two nodes. For the purposes of this solver, we will define unconnected graphs to have infinite weight. 
\begin{description}
\item[Monotonic Predicate:] $minSpanningTree_{G\leq C}(edges)$, true iff the minimum weight spanning tree in $G$ has weight $\leq C$.
\item[Algorithm:] Kruskal's algorithm~\cite{kruskal1956shortest} for minimum spanning trees.
\item[Conflict set for $minSpanningTree_{G\leq C}(edges)$:] The conflict set is $\{\lnot e_1, \lnot e_2,  \ldots,\\ minSpanningTree_{G\leq C}(edges)\}$, where $e_1, e_2, \ldots$ are the edges in some minimum spanning tree in $\gunder$.
\item[Conflict set for $\lnot minSpanningTree_{G\leq C}(edges)$:]~\\There are two cases to consider:
\begin{enumerate}
\item If $\gover$ is disconnected then we consider its weight to be infinite. In this case, we find a  cut $\{e_1, e_2,\ldots\}$ of disabled edges separating any one component from the remaining components. Kruskal's algorithm conveniently computes the components for disconnected graphs in the form of a minimum spanning forest. Given a component in $\gover$, a valid separating cut consists of all disabled edges $(u,v)$ such that $u$ is in the component and $v$ is not. We can either return the first such cut we find, or the smallest one from among all the components. For disconnected $\gover$, the conflict set is  $\{e_1, e_2, \ldots, \lnot minSpanningTree_{G\leq C}(edges)\}$.

\item If $\gover$ is \textit{not} disconnected, then we search for a minimal set of edges required to decrease the weight of the minimum spanning tree. To do so, we visit each disabled edge $(u,v)$, and then examine the cycle that would be created if we were to add $(u,v)$ into the minimum spanning tree. (Because the minimum spanning tree reaches all nodes, and reaches them exactly once, adding a new edge between any two nodes will create a unique cycle in the tree.) If any edge in that cycle has higher weight than the disabled edge, then if that disabled edge were to be enabled in the graph, we would be able to create a smaller spanning tree by replacing that larger edge in the cycle with the disabled edge. Let $e_1, e_2, \ldots$ be the set of such disabled edges that can be used to create lower weight spanning trees; the conflict set is $\{e_1, e_2, \ldots, \lnot minSpanningTree_{G\leq C}(edges)\}$.

In practice, we can visit each such cycle in linear time by using Tarjan's off-line lowest common ancestors algorithm~\cite{gabow1985linear} in the minimum spanning tree found in $\gover$. Visiting each edge in the cycle to check if it is larger than each disabled edge takes linear time in the number of edges in the tree for each cycle. Since the number of edges in the tree is 1 less than the number of nodes, the total runtime is then $O(|V|^2\cdot |D|)$, where $D$ is the set of disabled edges in $M$.
\end{enumerate}
\end{description}

\subsection{{Edges in Minimum Spanning Trees}}
Here, we consider a variation of the theory of minimum spanning tree weights.
Assuming that the minimum spanning tree for $G$ is unique (a
condition we can ensure by requiring that all the edges have unique
weights), 
we want to detect whether a particular edge is a member of the minimum spanning tree of $G$. 

Given an edge $(u,v)$ in $G$, if that edge is already in the minimum
spanning tree for $G$, then adding additional edges to $G$ can replace
$(u,v)$ in the minimum spanning tree. Conversely, if $(u,v)$ is in $G$
but is \textit{not} in the minimum spanning tree, then adding additional
edges cannot result in $(u,v)$ becoming part of the minimum spanning
tree.
We can define the predicate $edgeInTree_{u,v,G}(edges)$,
which is true iff the edge $(u,v)$ is in the minimum spanning tree of $G$, under assignment to the edges.

However, predicate $edgeInTree_{u,v,G}(edges)$ is not quite monotonic. If an edge is not enabled in the graph at all, then it isn't in the minimum spanning tree. Enabling the edge may result in a lower minimum spanning tree, putting $(u,v)$ into the tree. As we just established, it is then possible to remove $(u,v)$ from the minimum spanning tree by adding additional edges (after which point adding even more edges cannot return $(u,v)$ to the tree). Since  $edgeInTree_{u,v,G}(edges)$ can go from false to true and back to false by adding edges to $G$, it is non-monotonic with respect to its arguments. Instead, we propose a related predicate: $edgeInTreeOrDisabled_{u,v,G}(edges)$. This property is \textit{true} if the edge is disabled in $G$, \textit{or} if it is enabled and in the unique minimum spanning tree of $G$. We can then recover the original predicate in the SAT solver (outside of our theory solver), by setting $edgeInTree_{u,v,G}(edges) = edgeInTreeOrDisabled_{u,v,G}(edges) \land edge_{u,v,G}$. Implementing clause learning for this theory is very similar to the general MST solver above; we omit the details here due to space constraints.


\begin{description}
\item[Monotonic Predicate:] $edgeInTreeOrDisabled_{u,v,G}(edges)$.
\item[Algorithm:] Kruskal's algorithm~\cite{kruskal1956shortest} for minimum spanning trees.
\item[Conflict set for $edgeInTreeOrDisabled_{u,v,G}(edges)$:] If edge $(u,v)$ is disabled, the conflict set is $\{edge_{u,v,G}, edgeInTreeOrDisabled_{u,v,G}(edges)\}$. 
Otherwise, edge $(u,v)$ is enabled. Visit each cycle that would be formed by adding a disabled edge $e_i$ into the tree. If that cycle also includes edge $(u,v)$, and if any edge in the cycle has a higher weight than the disabled edge $e_i$, it might be possible to remove $(u,v$) from the tree by enabling edge $e_i$. The conflict set is $\{e_1, e_2, \ldots, edgeInTreeOrDisabled_{u,v,G}(edges)\}$.

\item[Conflict set for $\lnot edgeInTreeOrDisabled_{u,v,G}(edges)$:] The reason that $(u,v)$ is not in the minimum spanning tree of $\gunder$ is that the cycle formed by adding $(u,v)$ into the tree contains no edges that are higher weight than $(u,v)$ (or any that are equal, because we are assuming unique edge weights). We have from the cycle property of minimum spanning trees that $(u,v)$ cannot be in the tree unless one of the edges in that cycle is removed. We can find that cycle in linear time by walking up the tree from $u$ and $v$ to the lowest common ancestor of $u$ and $v$, and collecting each visited edge. Let $e_1, e_2,\ldots$ be the edges in the cycle. The conflict set is  $\{e_1, e_2,\ldots, \lnot edge(u,v,G),\lnot edgeInTreeOrDisabled_{u,v,G}(edges)\}$.
\end{description}

~\\
~\\
Having explored theories for several interesting graph properties, we now
consider a wider domain of monotonic theories: constraint satisfaction
and constrained optimization problems. Most constrained optimization
problems are monotonic with respect to the constraints in the problem, in
the sense that removing constraints can increase the optimum value of the
objective function, but cannot decrease it.\footnote{There \textit{are}
non-monotonic constrained optimization and satisfaction problems, but
they aren't common. Answer Set Programming and other default logics are
examples.}  Given a CSP formula $P$ with a finite set of constraints
$C$ (expressed in any monotonic propositional logic), we can support
predicates of the form $\textit{constraintEnforced}_{P}(constraint_i)$, and the monotonic
predicate $\textit{satisfiable}_P(enforcedConstraints)$. For optimization problems, we instead have
the predicate $\textit{optimumSolution}_{P \leq O}(enforcedConstraints)$, which is true iff the optimum
value of the objective function is $\leq$ $O$ under assignment to the enforced constraints.
This allows us to reason about which constraints are enforced in the
problem, and the satisfiability of those enforced constraints, but does
\textit{not} expose the actual solution of the CSP to the SAT solver.
For example, armed with a theory solver for the traveling salesman problem,
an SMT solver could reason about which cities must be visited in the tour,
and whether or not a tour can be found of less than a certain length,
but not reason about the path of the optimum tour itself. To illustrate this general construction for constraint
satisfaction problems, we present a simple scheduling theory.



\subsection{{Preemptive Scheduling for Uniprocessors}}
Consider the constraint satisfaction problem of preemptive scheduling on a uniprocessor. This can be solved optimally in polynomial time using the earliest deadline first scheduling policy\cite{liu1973scheduling}. Here we consider atoms of the form $task_{A,L,D,P}$, where $A$ is the arrival time of the task, $L$ is the required time for the task, $D$ is the deadline by which time the task must be completed, and $P$ is the processor to schedule the task on. A $task_{A,L,D,P}$ must be scheduled on processor $P$ if the atom is true in $M$; if it is false, then the task does not need to be scheduled at all. The theory's monotonic predicate is $schedulable_P(tasks)$, where $tasks$ is a vector of Boolean variables corresponding exposed by the $task_{A,L,D,P}$ atoms, which is true iff the tasks assigned to $P$ can be scheduled. The theory can easily be extended to support the predicate $schedulableInTime_{P \leq T}(tasks)$, which is true if the tasks can be scheduled in $T$ or fewer time slots, if desired. 

\begin{description}
\item[Monotonic Predicate:] $schedulable_P(tasks)$.
\item[Algorithm:] Earliest deadline first (EDF).
\item[Conflict set for $schedulable_P(tasks)$:] Let $task_1, task_2,\ldots$ be the disabled tasks of $P$ in $M$; the conflict set is $\{task_1, task_2, \ldots, schedulable_P(tasks)\}$
\item[Conflict set for $\lnot schedulable_P(tasks)$:] Let $task_1, \ldots, task_k$ be the tasks successfully scheduled by EDF, with $task_k+1$ the first unschedulable task found. 
Let $task_i$ be the last task scheduled by EDF that starts at its arrival time (at least one such task, $task_1$, is guaranteed to exist); the conflict set is $\{\lnot task_i, \lnot task_{i+1} \ldots, \lnot task_{k+1}, \lnot schedulable_P(tasks) \}$.

\end{description}

\section{Applications and Results \label{sec:graph}}

Many popular video games --- including independent titles like \textit{Dwarf Fortress} and \textit{Minecraft}, and also mainstream games such as the\textit{ Elder Scrolls} series --- include procedurally generated content. All the games just mentioned prominently feature large, rolling landscapes that were partially or fully generated programmatically; Dwarf Fortress also generates complex historical records for its setting. Recently, there has been interest in \textit{declarative} procedural content generation (e.g.,~\cite{boenn2008automatic,smith2011answer}), in which the artifact to be generated is specified as a solution to a logic formula.

Many procedural content generation tasks are really graph generation
tasks; for example, in maze generation, the goal is to select a set
of edges to include in a graph  (from some set of possible edges that
may or may not form a complete graph)  such that there exists a path
from the start to the finish, while also ensuring that when the graph
is laid out in a grid, the path is non-obvious.
For example, the open-source terrain
generation tool Diorama\footnote{http://warzone2100.org.uk} considers
a set of undirected, planar edges arranged in a grid. Each position on
the grid is associated with a height; Diorama searches for a heightmap
that realises a complex combination of desirable characteristics
of this terrain, such as the positions of mountains, water, cliffs, and
player's bases, while also ensuring that all positions in the map are reachable
from some starting point. Edges from this grid are only included in the
graph if the heightmap does not have a sharp elevation change (a cliff)
between the endpoints of the edge.  Diorama expresses its constraints
in Answer Set Programming (ASP)~\cite{baral2003knowledge} --- a logic
formalism closely related to SAT, and with efficient
solvers~\cite{gebser2007clasp} based on state-of-the-art
CDCL SAT solvers. Unlike SAT, ASP can encode reachability constraints in
cyclic graphs in linear space, and ASP solvers can solve the resulting
formulae efficiently in practice. Partly for this reason, ASP solvers
are more commonly used than SAT solvers in declarative procedural content
generation applications.   For instance,
Diorama, Refraction~\cite{smith2011answer},
and Variations Forever~\cite{smith2010variations} all use ASP.

For each of the theories in the preceding section, we provide comparisons of our SMT solver \monosat (based on \minisat) against the state-of-the-art ASP solver \clasp 2.14 (and, where practical, also to \minisat 2.2) on realistic procedural content generation problems. All experiments were conducted on an Intel i7-2600K CPU, at 3.4 GHz (8MB L3 cache), limited to 900 seconds and 16 GB of RAM. Reported runtimes for \clasp{} do not include the cost of grounding (which varies between instantaneous and hundreds of seconds, but in procedural content generation applications is typically a sunk cost that can be amortized over many runs of the solver).

Since these experiments were conducted, a new version of \clasp (4.3.0) has been released; additionally, in consultation with an ASP expert, we have found that some of our ASP encodings are sub-optimal; in particular, the connected components results have been dramatically improved upon for \clasp. Results for our own solver for reachability and shortest paths have been greatly improved and will be reported in the final version of this study.


\paragraph{\textbf{Reachability:}}
For acyclic graphs, reachability can be encoded in CNF in linear space and solved efficiently. However, the standard SAT encoding for general graphs is quadratic in the number of vertices and quickly becomes too expensive to be practical, even for small graphs. In contrast, reachability constraints can be encoded in ASP in $O(|V| + |E|)$ space.  
We include our graph reachability solver for completeness; \clasp can already solve these constraints very efficiently (and, as we will see, typically more efficiently than our solver can). 

We consider two applications for the theory of graph reachability. The first is a subset of the \textit{cliff-layout} constraints from the terrain generator Diorama.\footnote{Because we had to manually translate these constraints from ASP into our SMT format, and because our solver doesn't support pseudo-boolean constraints, we use only a subset of these cliff-layout constraints. Specifically, we support the \textbf{undulate}, \textbf{sunkenBase}, \textbf{geographicFeatures}, and \textbf{everythingReachable} options from cliff-layout, with near=1, depth=5, and 2 bases (except where otherwise noted).} 
\LNCSFigureSquareTwo{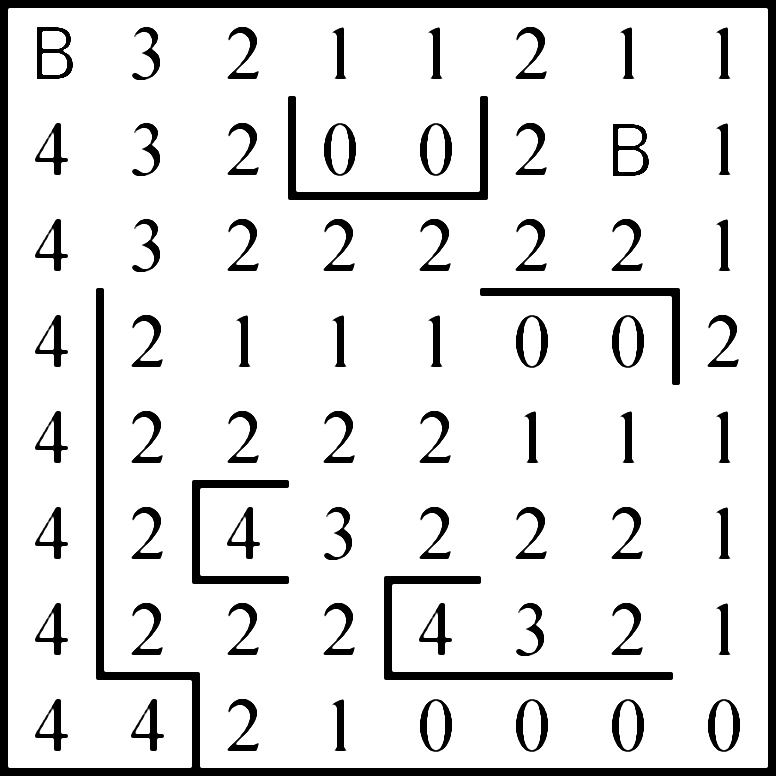}{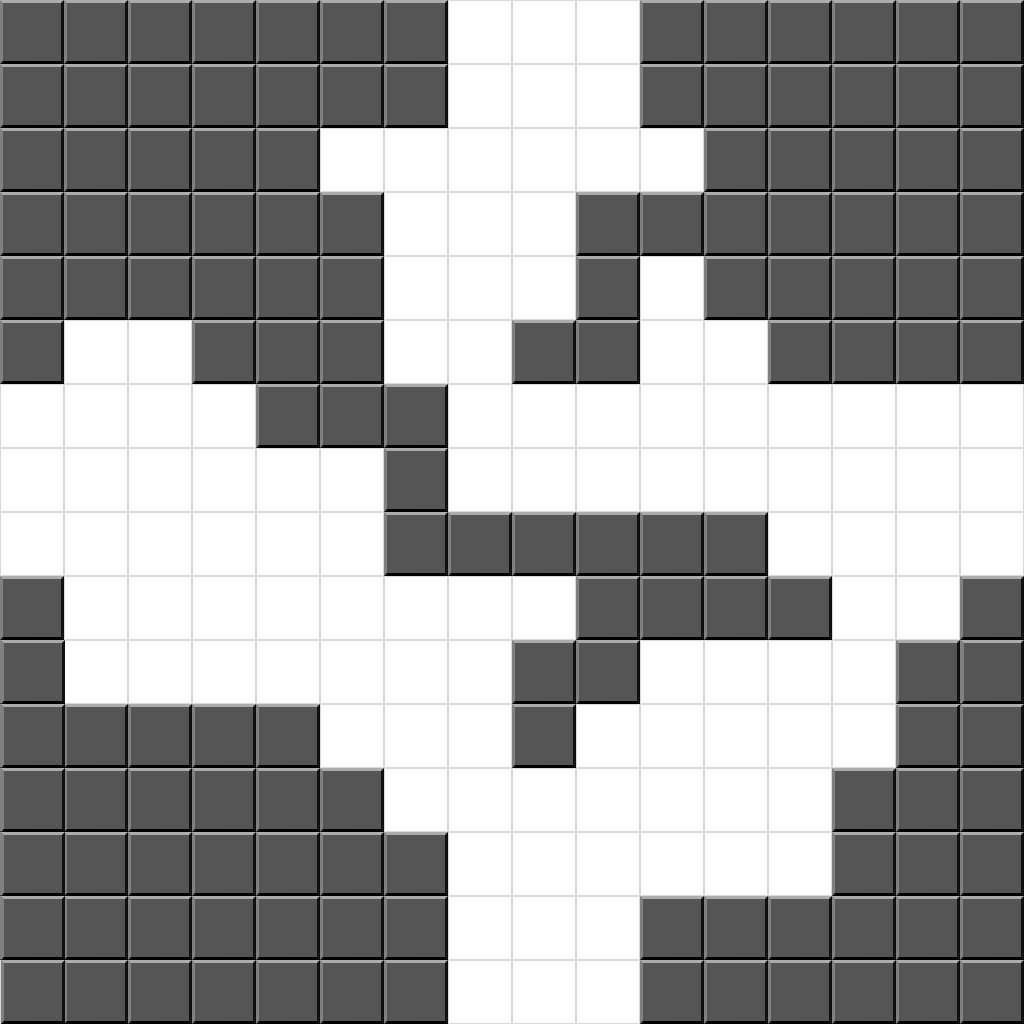}{Generated terrain}{Left, a heightmap generated by Diorama, and right, a cave (seen from the side, with gravity pointing to the bottom of the page) generated in the style of 2D platformers. Numbers in the heightmap correspond to elevations (bases are marked as `B'), with a difference greater than one between adjacent blocks creating an impassable cliff. In the platformer, players must traverse the room by walking and jumping --- complex movement dynamics that are modeled as edges in a graph.}

\begin{wraptable}{O}{0.6\textwidth}

\centering
\begin{tabular}{ l c c c c c }
  Reachability & \monosat &  \clasp & \minisat \\
  \hline
   Diorama 16x16  & 12s &   $< 0.01$s &  Timeout \\ 
   Diorama 32x32  & Timeout & 0.2s  & n/a \\       %
   Platformer 16x16 & 0.8s & 1.5s & Timeout \\
   Platformer 24x24 & 277s  & Timeout & n/a\\
  \hline
\end{tabular}
\comments{
\begin{tabular}{ l c c c c c }
  Solver & Encoding  &  Diorama 16x16 & 32x32 & Platformer 16x16 & 24x24\\
  \hline
  \minisat & $\mathcal{O}(|E|^2)$ & Timeout  & -- & Timeout & --\\
  \clasp &  $\mathcal{O}(|E|)$ &  $< 0.01$ s & 0.2 s  & 1.5s & Timeout \\
  \monosat &  $\mathcal{O}(|E|)$  & 3s & 237  & 0.8s & 277s\\
  \hline
\end{tabular}
}
\caption{Reachability Results.\label{table:reach}
Notice: These results are preliminary.
}
\end{wraptable}
The second example is drawn from a 2D sidescrolling videogame. This
game\footnote{Developed by the first two authors with their
brothers David and Jacob Bayless.} generates rooms in the style of
traditional \textit{Metroidvania} platformers. Reachability
constraints are used in two ways: first, to ensure that the air and
ground blocks in the map are contiguous, and secondly, to ensure that
the player's on-screen character is capable of reaching each exit
from any reachable position on the map. This ensures not only that there
are no unreachable exits, but also that there are no traps in the room.

Runtime results in 
Table~\ref{table:reach} show that while our solver performs
much better than plain SAT (for which the larger instances aren't
even practical to encode, indicated as `n/a' in the table),  \clasp is much faster than our solver on the
Diorama instances. On the other hand, our solver outperforms \clasp on
the platformer constraints --- but only when we use our optional decision
heuristic for reachability queries (without it, we time out; on the
Diorama instances, it has no impact on runtime). In fact, there is good
reason to expect that \clasp would outperform our solver on typical
reachability queries:  \clasp not only supports a linear encoding
for reachability constraints,
but is well-optimized (and widely used) for these types of reachability
constraints, including the ability
to apply incremental
unit propagation on the constraints. In contrast, our solver must
repeatedly recompute the reachable nodes in $\gunder$ and $\gover$, from
scratch, after each decision. Even if we were to employ a fully dynamic
reachability algorithm, at each decision, we would still be doing roughly twice as much work
(updating both the over and under-approximation)  as
\clasp{}. The only case where we would expect to see an advantage
over \clasp on reachability would be if there are special properties
of the graph  that we can exploit (as
in the case of our decision heuristic).
In contrast, for all the other monotonic theories we present,
the encoding into ASP (and SAT) is non-linear, and in each case we will outperform \clasp dramatically.

\paragraph{\textbf{Shortest Paths:}}

\begin{wraptable}{R}{0.45\textwidth}

\centering
\begin{tabular}{ l c c c c c }
  Shortest Paths & \monosat &  \clasp  \\
  \hline
   Diorama 8x8 & 11s & 40s  \\ 
   Diorama 16x16 &  130s   &  508s \\ 
  \hline  
\end{tabular}~~~~
\caption{Shortest Paths Results. \label{table:shortest}Notice: These results are preliminary.}
\end{wraptable}

We considered a modified version of the Diorama terrain generator, replacing the reachability constraint with the constraint that the distance (as the cat runs, not as the crow flies) between any two bases must be between 15 and 35 steps (for 8 by 8 maps) or between 25 and 45 steps (for 16 by 16 maps). We tried this for two map sizes, 8 by 8, and 16 by 16. 
\comments{
Our second application is the chromatic maze generation problem from~\cite{smith2011answer}. This task involves assigning colors to a grid to form a `chromatic' maze, with a minimum start-to-finish path length of a certain size (for the 6x6 maze, the path must be at least 20 steps; for the 10x10 maze, at least 65 steps). 
}

Like reachability, the standard encoding for shortest paths into SAT is
quadratic in the number of nodes of the graph; unlike reachability, the standard
encoding into ASP is also quadratic. Table~\ref{table:shortest}
shows large performance improvements over \clasp (we also ran experiments with \minisat, which
timed out on all instances). \comments{Smith~\cite{smith2011answer}
reported
that \clasp required two hours to find a 114-step maze
in a 21x21 map, with even longer mazes requiring more time. In
contrast, our solver can generate a 114 step maze in just 7 seconds. We
can produce larger mazes as well: a 161-step maze required 26 seconds,
and a 221-step maze in 156 seconds.\footnote{These mazes were verified
by appending their solutions to the original chromatic maze ASP generator
from~\cite{smith2011answer} and solving with \clasp.}
}

\paragraph{\textbf{Connected Components:}}

\begin{wraptable}{R}{0.45\textwidth}
\centering
\begin{tabular}{ l c c c c c }
  Components & \monosat &  \clasp  \\
  \hline
   8 Components &  6s  & 98s\\
    10 Components & 6s &  Timeout \\  
    12 Components  & 4s & Timeout \\
    14 Components & 0.82s &   Timeout  \\
    16 Components & 0.2s &   Timeout  \\
   
  \hline  
\end{tabular}
\caption{Connected Components Results.\label{table:component} 
Notice: These results have been obsoleted by a greatly improved encoding for \clasp.
}
\end{wraptable}

We modify the Diorama constraints such that the generated map must consist of exactly 10 different terrain `regions', where a region is a set of contiguous terrain positions of the same height. This produces terrain with a small number of large, natural-looking, contiguous ocean, plains, hills, and mountain regions.

For this problem, we disabled the `undulate' constraint, as well as the reachability constraint (as neither \clasp nor our solver could solve the problem with these constraints combined with the connected components constraint). Results are presented in Table \ref{table:component}.


\paragraph{\textbf{Maximum-Flow/Minimum-Cut:}}
We modify the Diorama constraints such that each edge has capacity 1, and use the theory of maximum-flows to enforce that the minimum cut between the top and bottom of the map must not be less than 4. This prevents chokepoints between the top and bottom of the map. However, this constraint tends to produce artificial-looking passages and long, straight corridors in the map. For this reason, we consider a second variation, in which the edges have random positive integer capacities between 1 and 4, and the minimum cut between the top and bottom of the map must not be less than 8 (doubled to account for the increase in the average capacity of the edges). This again avoids chokepoints, but produces more natural looking cliffs and canyons with undulating widths.

\begin{wraptable}{R}{0.45\textwidth}

\begin{tabular}{ l c c c c c }
  Maximum Flow & \monosat &  \clasp  \\
  \hline
   Capacity 1 & 14s & 22s  \\
   Capacity 1 to 4 &  21s   &  168s \\
   Capacity 4 & 14s &   Timeout \\
  \hline  
\end{tabular}~~~~
\caption{Maximum Flow Results. $F$ is the maximum $s-t$ flow in $G$.\label{table:MST}. Notice: These results are preliminary.}
\end{wraptable}

In Table \ref{table:mincut}, we show that when the edge capacities are 1, \clasp performs almost as well as \monosat, but that with larger network flows \monosat is much faster than \clasp. For example, if we simply multiply the edge capacities and chokepoint constraint by 4 (third row of Table \ref{table:mincut}), then \clasp times out, even though the constraints are logically equivalent to the (easy) first example we considered. This is a direct consequence of the cost of encoding the range of possible flows along each edge in ASP.

\paragraph{\textbf{Minimum Spanning Trees:}}

\begin{wraptable}{R}{0.5\textwidth}
\vspace*{-3ex}
\begin{tabular}{ l c c c c c }
  Spanning Tree & \monosat &  \clasp  \\
  \hline
  Maze  5x5  & 0.01s &    15s  \\
  Maze  8x8  & 1.5s &   Timeout  \\
  Maze  16x16   & 32s & Timeout \\
  \hline  
\end{tabular}
\caption{Minimum Spanning Tree Results. $C$ is the maximum weight of any edge.\label{table:mincut} Notice: These results are preliminary.}

\comments{
\caption{Minimum Spanning Tree Weights and Edges. $C$ is the maximum weight of any individual edge.\label{table:MST}}
\centering
\begin{tabular}{ l c c c c c c}
  Solver & Encoding  &  Maze Generation  5x5  & 8x8 & 16x16 \\
  \hline
  \clasp &  $\mathcal{O}(|E|C^2)$ & 15s  & Timeout & Timeout  \\
  \monosat &  $\mathcal{O}(|E|)$ & 0.01s & 1.5s & 32s\\
\hline
\end{tabular}
}

\comments{
\centering
\begin{tabular}{ l c c c c c c}
  Solver & Encoding  &  Diorama 16x16 & Random Weights & Large Weights.  \\
  \hline
  \clasp &  $\mathcal{O}(|V|^2|E| + |E|F^2)$ & 10s & 30s &  Timeout   \\
  \monosat &  $\mathcal{O}(|E|)$ & 3s & 3s & 3s\\
\hline
\end{tabular}
}
\end{wraptable}

A common approach to generating random, traditional, 2D pen-and-paper mazes, is to find the minimum spanning tree of a randomly weighted graph. Here, we consider a related problem: generating a random maze with a shortest start-to-finish path of a certain length.


We model this problem with two graphs, $G_1$ and $G_2$. In the first graph, we have randomly weighted edges arranged in a grid. In the second graph we have all the same edges, but unweighted. Edges in $G_2$ are constrained to be enabled if and only if the corresponding edges in $G_1$ are elements of the minimum spanning tree of $G_1$. We then enforce that the shortest path in $G_2$ between the start and end nodes is between 3 and 4 times the width of the graph. Since the only edges enabled in $G_2$ are the edges of the minimum spanning tree of $G_1$, this condition constrains that the path length between the start and end node in the minimum spanning tree of $G_1$ be within these bounds. Finally, we constrain the graph to be connected (we could use the connected components theory above, but as we are already computing the minimum spanning tree for this graph in the solver, we can more efficiently just enforce the constraint that the minimum spanning tree has weight less than infinity).

\LNCSFigureSquareTwo{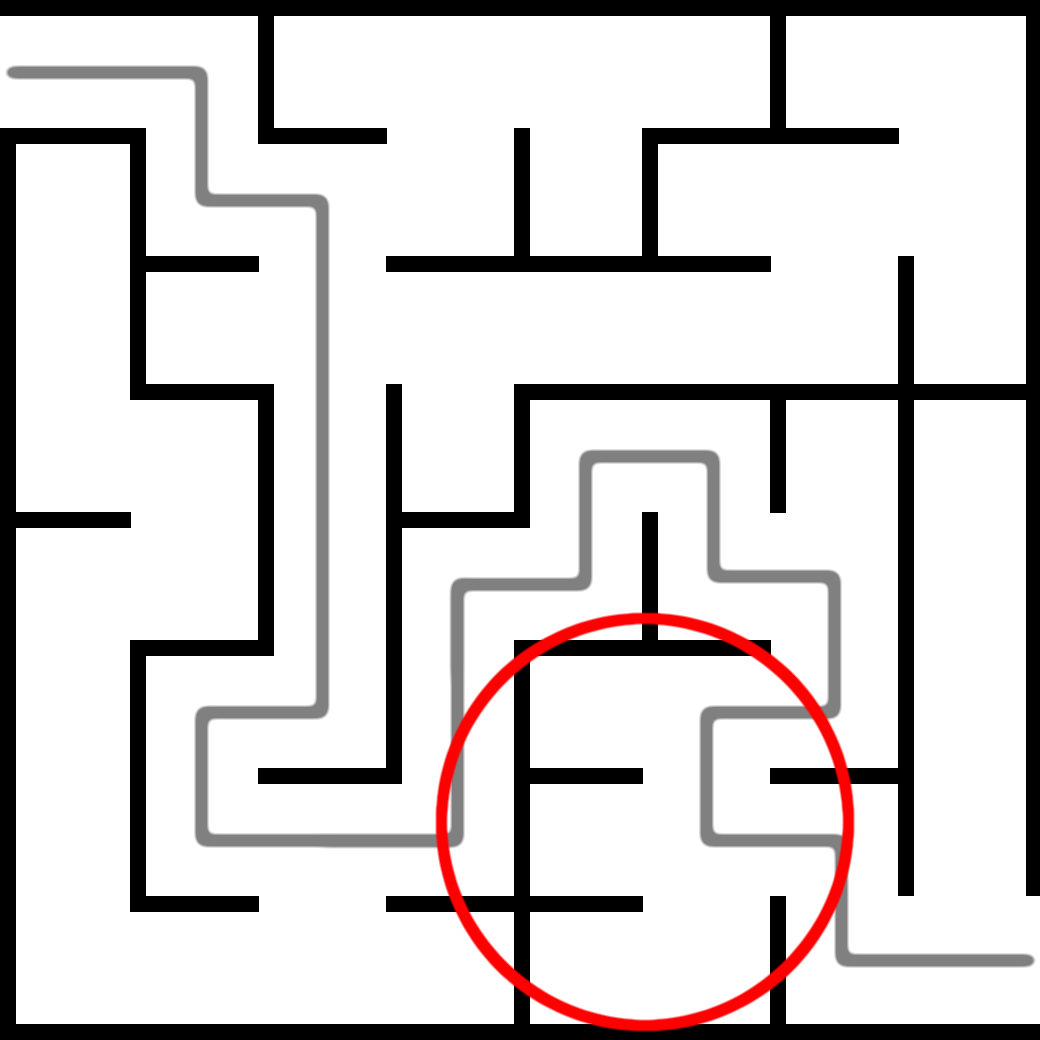}{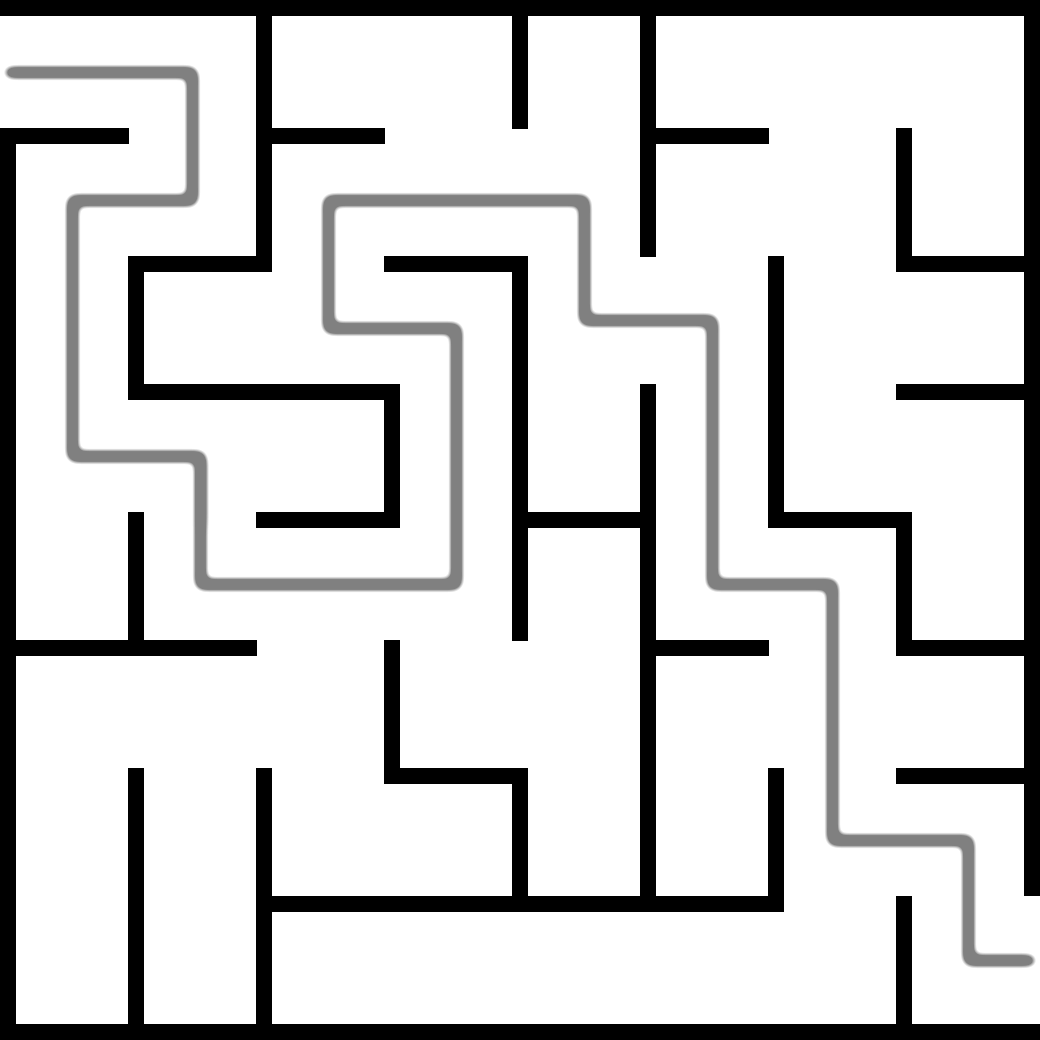}{Random Mazes}{Mazes generated through a combination of minimum spanning tree edge and weight constraints, and a constraint on the length of the path from start to finish. On the left, an un-optimized maze, with awkward, comb-structured walls (circled). On the right, an optimized maze, generated in seconds by \monosat.\label{mazes}}

The solver must then select a set of edges in $G_1$ to enable and disable such that the minimum spanning tree of the resulting graph is a) connected and b) results in a maze with a shortest start-to-finish path within the requested bounds. By itself, these constraints can result in poor-quality mazes (see Figure \ref{mazes}, left, and notice the unnatural wall formations near the bottom right corner); by allowing the edges in $G_1$ to be enabled or disabled freely, any tree can become the minimum spanning tree, effectively eliminating the effect of the random edge weight constraints.

Instead we convert this into an optimization problem, by combining it with an additional constraint: that the minimum spanning tree of $G_1$ must be $\leq$ to some constant, which we then lower repeatedly until it cannot be lowered any further without making the instance unsatisfiable (\clasp supports this operation via the ``\#minimize'' statement). This produces plausible mazes (see Figure \ref{mazes}, right), while satisfying our shortest path constraints.

\paragraph{\textbf{Preemptive Scheduling for Uniprocessors:}}
Preemptive uniprocessor scheduling, by itself, isn't a very interesting
theory, but we demonstrate how much more expressive it can be when
combined with general Boolean reasoning.
Our example is a randomized scheduling problem with 1000 tasks and
a parameter \textit{slack}.
Each $task_i$ has a random duration of 1 to 5 seconds,
an arrival time in the range $(0,1000-\textit{slack})$, and a
deadline at the task's arrival time + $\textit{slack}$.
We model multiprocessor scheduling
by creating 10 separate uniprocessor theory solvers $P_0 \ldots P_9$.
The processors are heterogenous, with 
each processor $P$ having a slowdown factor $1 \le s_P \le 2$.
Accordingly, for each $task_i$, 
we instantiate theory atoms $task(A_i, s_P \cdot  L_i, D_i, P_p)$ for each processor $P_p$.
We enforce via SAT constraints
that each $task_i$ be scheduled on exactly one processor. 
We further
\begin{wraptable}{r}{0.45\textwidth}

\centering
\begin{tabular}{ l c c c c c }
  Scheduling & \monosat &  \clasp  \\
  \hline
  Slack=10     & 73s & 30s  \\
    Slack = 25  & 70s &    Timeout  \\
  Slack = 50  & 38s &    Timeout  \\
  Slack = 100 & 47s & Timeout \\

  \hline  
\end{tabular}
\comments{
\begin{tabular}{ l c c c c c }
  Scheduling & \monosat &  \clasp  \\
  \hline
  Encoding   & $\mathcal{O}(J)$ &  $\mathcal{O}(J \cdot T \cdot L)$  \\  
  50  Tasks   & $<0.01s$ & Timeout  \\
  100 Tasks  & 0.01s &    Timeout  \\
  1000 Tasks & 20s & Timeout \\

  \hline  
\end{tabular}
}
\caption{Preemptive Scheduling for Uniprocessors. $J$ is the number of jobs; and $S$ is $\max{slack}$ of any task.\label{table:sched} Notice: These results are preliminary.}
\comments{
\begin{tabular}{ l c c c c c c}
  Solver & Encoding  &  \#Jobs  100  & 500 & 1000 \\
  \hline
  \clasp &  $\mathcal{O}(JT^2)$ & 0.2s  & Timeout & Timeout  \\
  \monosat &  $\mathcal{O}(J)$ & 0.01s & 2s & 5s \\
\hline
\end{tabular}
}
\vspace*{-4ex}
\end{wraptable}
extend our example to model transactions (groups
of tasks that must be scheduled in an all-or-none manner), by randomly
partitioning the tasks into groups of 10, and enforcing that the jobs in
each partitioned are either all scheduled (not necessarily on the same
processor), or not scheduled at all.  Finally, we enforce that exactly
half of these tasks are successfully scheduled.

Table~\ref{table:sched} shows that for very small slack times ($slack=10$),
\clasp is faster than \monosat, but as the slack grows to even moderate sizes,
the ASP encoding quickly becomes impractical to solve, while \monosat scales efficiently.

\comments{
, of
which the SAT solver must choose half to be scheduled within 50, 100,
or 1000 time slots, respectively. 
As can be seen in
Table \ref{table:sched}, our theory solver scales well as the problem
size increases.
}


\vspace*{-1ex}
\section{Conclusion}

We have introduced the concept of a monotonic theory
and showed a systematic technique to build efficient SMT solvers incorporating
such theories.  Our technique leverages common-place, highly efficient
algorithms for fully specified problem instances, in order to achieve
efficient theory propagation and clause learning from the partially specified
instances that arise in a lazy SMT approach.
We demonstrate the generality of the monotonic theory concept by providing
several example theories drawn from graph theory, and one theory to illustrate
how typical constrained optimization problems yield monotonic theories.
These example theories are expressive --- permitting compact encodings for
real problems arising from procedural content generation and scheduling ---
and the SMT solvers we produce via our technique (and standard,
unmodified graph and scheduling algorithms) perform well in practice.

As mentioned earlier, fully dynamic algorithms
are available for all graph properties considered here
(e.g.,~\cite{henzinger1995fully,thorup1999undirected,thorup2000near,holm2001poly,thorup2001fully}).
These dynamic algorithms permit efficient recomputation of graph
properties as edges are added to and removed from a graph, without
having to start from scratch each time. So far, we have not explored this
direction, 
but it is a clear avenue for future improvement.
The most immediate direction for future work, however, is
to discover additional
monotonic theories and new application domains that can benefit from them.


\newpage

\bibliographystyle{plain}
\bibliography{smmt}

\end{document}